\documentclass[fleqn,twoside]{article}
\usepackage{espcrc2}      
\usepackage{psfig}

\title{Doping dependence of the electronic Raman spectra in cuprates}  
\author{F. Venturini\address[WMI]{Walther Meissner Institute, Bavarian
Academy of Sciences, D-85748 Garching, Germany},
M. Opel\addressmark[WMI], R. Hackl\addressmark[WMI], 
H. Berger\address[EPFL]{EPFL, Ecublens, CH-1015 Lausanne, Switzerland},
L. Forr\'o\addressmark[EPFL]
 and B. Revaz\address{DPMC, University of Geneva, CH-1211 Geneva,
Switzerland}} 

\begin{document}

\begin{abstract}
We report electronic Raman scattering measurements on 
Bi$_2$Sr$_2$(Y$_{1-x}$Ca$_x$)Cu$_2$O$_{8+\delta}$  single crystals at
different doping levels. The dependence of the spectra on doping and
on incoming photon energy is analyzed for different polarization
geometries, in the superconducting and in the normal state.
We find the scaling behavior of the superconductivity pair-breaking
peak with the carrier concentration to be very different in B$_{1g}$
and B$_{2g}$ geometries. 
Also, we do not find evidence of any significant variation of the
lineshape of the spectra in the overdoped region in both symmetries,
while we observe a reduction of the intensity in B$_{2g}$ upon
decreasing photon energies. 
The normal state data are analyzed in terms of the memory-function
approach. 
The quasiparticle relaxation rates in
the two symmetries display a dependence on energy and temperature
which varies with the doping level.
\end{abstract}

\maketitle

\section{Introduction}

Inelastic light-scattering (Raman) experiments reveal a
two-particle response of interacting electrons in a similar way
as infrared
spectroscopy hence combining advantages of angle resolved
photoemission spectroscopy (ARPES), by allowing momentum resolution,
and of conductivity measurements.
Raman studies
of the electron properties in cuprates have already been carried 
out for various doping levels in samples of
Bi$_2$Sr$_2$(Y$_{1-x}$Ca$_x$)Cu$_2$O$_{8+\delta}$ (Bi-2212),
YBa$_2$Cu$_3$O$_{6+x}$ (Y-123), 
and La$_{2-x}$Sr$_x$CuO$_4$ (LSCO) \cite{Kendziora95,Chen,Opel00,Sugai00}.
However, there are still many unclarified issues as well as
considerable improvements in sample quality which motivate a detailed
analysis of electronic properties below and above $T_c$ with doping
$p$. In fact, for the understanding of superconductivity in the
cuprates the properties of the normal state are considered to
be as crucial as those below $T_c$ since the relevant interactions
have a characteristic influence on the carriers at all temperatures.

Through the choice of the incident and scattered polarization
vectors, Raman scattering is sensitive to different portions of the
Fermi surface. 
By selecting the incident and scattered light polarizations parallel
to $x$ and $y$ directions respectively, $x$ and $y$ being parallel to
the Cu-O bonds, the B$_{2g}$ symmetry is projected out. In this case the
excitations with momenta along the diagonals of the Brillouin zone
(BZ) are mainly 
probed. With polarizations along $x'y'$ (at 45$^{\circ}$ to the Cu-O
bonds) the B$_{1g}$ symmetry is selected. Then the
excitation probed are those with momenta along the BZ axes.

In this paper we report electronic Raman scattering measurements in
B$_{1g}$ and B$_{2g}$ polarization geometries at different doping
levels in the superconducting and in the normal 
state. For $T>T_c$ the data are analyzed in terms of the memory
function approach recently introduced to
extract dynamical relaxation rates of the carriers \cite{Opel00}.

\section{Experiment}

High quality Bi$_2$Sr$_2$(Y$_{1-x}$Ca$_x$)Cu$_2$O$_{8+\delta}$
single crystals were prepared in ZrO crucibles. They cover a wide doping
range as 
summarized in Table \ref{table1}. The doping level $p$ is
calculated from the relation  
$T_c=T_c^{max} [1-82.6(p-0.16)^2]$ \cite{Tall95}.

\begin{table}[hbt]
\renewcommand{\arraystretch}{1.2} % enlarge line spacing             
\caption{Bi-2212 single crystals studied.}
\begin{tabular}{l|ccccccc}
\hline
$T_c$(K) & 57 & 92 & 92 & 78 & 62 & 56\\
$p$ & 0.09 & 0.15 & 0.16 & 0.20 & 0.22 & 0.23 \\
\hline
\end{tabular}
\label{table1}
\end{table}

The experiments were performed in pseudo back-scattering geometry
using a standard Raman setup and an Ar$^+$ laser for excitation
at 458 nm, 514 nm and 528 nm, and a Kr$^+$ laser at 647 nm.
To reduce the heating of the sample due to the absorbed radiation, the
power of the laser was maintained below 4 mW.
The temperatures given below are corrected for the laser-induced
heating as estimated from the ratio of the Stokes to the anti-Stokes
intensities. 
The measured spectra are divided by the Bose-Einstein factor to
display the Raman response $\chi ''$.

\section{Superconducting state}

\begin{figure}[htb]
\centerline{\psfig{file=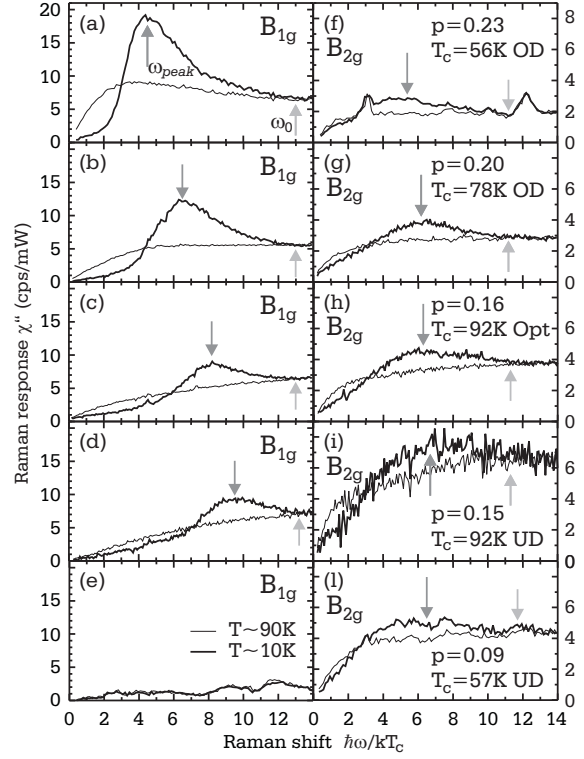,width=7.5cm}}
\caption{Doping dependence of the Raman spectra in B$_{1g}$ and
B$_{2g}$ symmetries in the superconducting and 
normal state. The arrows mark the $\omega_{peak}$ and $\omega_0$
energies.}
\label{doping}
\end{figure}

Fig.~\ref{doping} shows the B$_{1g}$ and B$_{2g}$ electronic Raman
spectra in differently doped Bi-2212 samples in the superconducting
state at approximately 10 K and in the normal state at approximately 90 K.
%In all the spectra the phonon contribution have been subtracted.
The data are plotted as a function of energy normalized to the
respective transition temperatures of the samples to better
visualize the scaling of relevant energies with $T_c$. 

The superconductivity-induced features in B$_{1g}$ and B$_{2g}$
symmetries show a very different scaling behavior with the carrier
concentration. 
We will discuss first the B$_{1g}$ symmetry results.
As the carrier concentration decreases from overdoped,
Fig.~\ref{doping}(a), to underdoped, Fig.~\ref{doping}(e), the
B$_{1g}$ spectrum undergoes substantial modifications. In particular,
the intensity is strongly suppressed when reducing the doping level,
and for  $p<0.15$ there is no evidence of
superconductivity-induced features (Fig.~\ref{doping}(e)), since there
is no observable difference between the superconducting and the normal
state spectra. 
The energy of the superconductivity-induced feature
$\omega_{peak}$, is also strongly 
affected by doping, going from $\hbar\omega_{peak} / k_BT_c \sim 4.5$
in the most overdoped sample (Fig.~\ref{doping}(a)) to 
$\hbar\omega_{peak} / k_BT_c \sim 8.7$ in the slightly underdoped 
one (Fig.~\ref{doping}(d)). 
From these observations we conclude that, in the doping range
studied, the peak energy is not proportional to $k_BT_c$ but decreases
monotonically with increasing $p$. 
It is worth noticing that we observe the same doping dependence in Y-123
samples \cite{Nem98}.

In contrast, the  B$_{2g}$ spectra show a superconductivity-induced
feature at all doping levels as it is evident from
Fig.~\ref{doping}(f) to Fig.~\ref{doping}(l). In addition, the   
maxima of the superconducting spectra $\hbar \omega_{peak} / k_BT_c$
are roughly constant between 6 and 6.5, with the exception of the the
most overdoped sample (Fig.~\ref{doping}(f)).

\begin{figure*}[htb]
\psfig{file=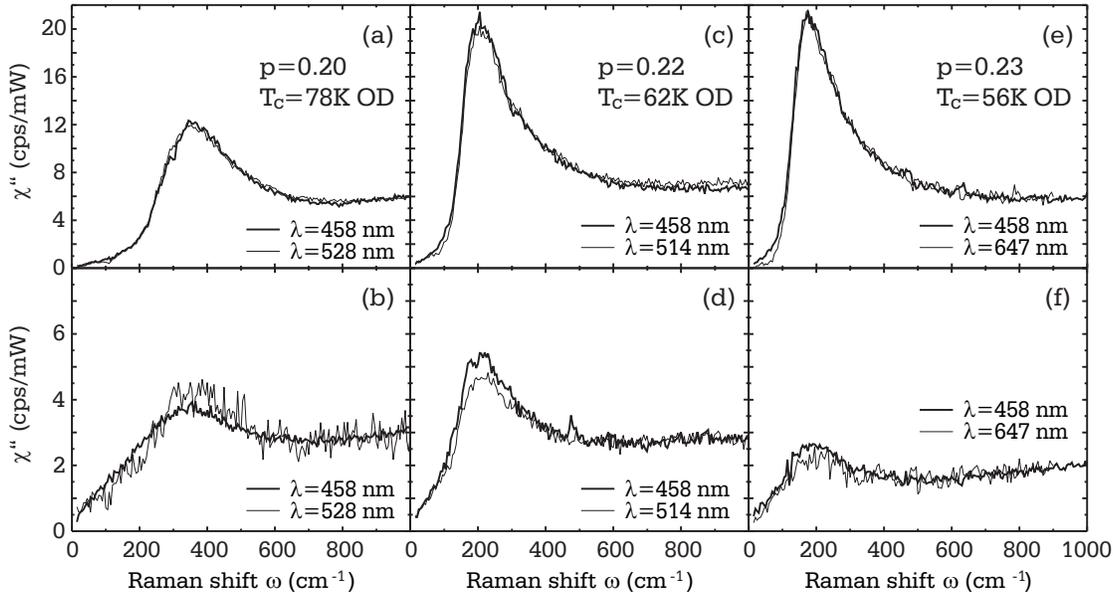}
\caption{B$_{1g}$ and Raman B$_{1g}$ spectra of overdoped Bi-2212
samples in the superconducting state ($T\sim 10$ K) at different
excitation energies as indicated.}  
\label{wavelength}
\end{figure*}

Assuming the superconducting gap to vary as
$\Delta({\bf k}) =\Delta_0/2 \ (\cos(k_x)-\cos(k_y))$, 
it is possible to extract the maximum $\Delta_0$ \cite{TPD}. 
In the B$_{2g}$ symmetry the measured spectra support $\Delta_0$
scaling with $T_c$. In addition, the effects of superconductivity are
observable up to an energy $\hbar\omega_0$ (see Fig.~\ref{doping}),
where the superconducting and normal spectra merge, which again scales
with $T_c$, both in  B$_{1g}$ and B$_{2g}$ symmetries, at any doping
level studied.

The scaling of $\Delta_0$ with $T_c$ is consistent with the results
for the superconducting energy gap derived from the magnetic
penetration depth \cite{Pana98}, electron tunneling for low bias
voltages well below the gap \cite{renner98}, as well as Andreev
reflection measurement \cite{Deut99}. 
However, if the B$_{1g}$ peak energy $\hbar \omega_{peak}$ is
considered, Raman scattering seems to reveal an additional energy
scale which increases monotonically when decreasing the doping
level. This can not be 
explained in the framework of a simple $d$-wave
superconducting energy gap, consistently with the B$_{2g}$ symmetry
and requires further theoretical work.
In addition, the doping dependence of $\omega_{peak}$
resembles that of the energy scale probed by ARPES and tunneling
experiments \cite{Miya98,Mesot99}. 

It has been suggested that there are two relevant energy scales,
namely the single-particle excitation energy, probed by 
ARPES and tunneling spectroscopies, which increases by decreasing the
doping, and a coherence energy scale, obtained for example by Andreev
reflections or B$_{2g}$ Raman scattering \cite{Deut99}. These
two energies approach the same value at high doping levels,
where the material is believed to recover a BCS-like behavior, and 
become increasingly different at low dopings.
The interpretation of these energy scales and their relation is still
an unsolved question.
However, to clearly address these questions more experiments are
required to investigate the $\omega_{peak}$ scaling behavior in the
underdoped side of the phase diagram and at which dopings the
pair-breaking peak disappears in the B$_{1g}$ symmetry.

In order to further investigate the superconductivity-induced features, 
we have studied the spectra of the three overdoped Bi-2212
samples at different excitation energies. 
In fact, it has been previously argued that resonance properties of
the pair-breaking excitations are observable in highly overdoped Bi-2212 
samples in B$_{1g}$ symmetry, and that they are possibly a signature of
an antiferromagnetically correlated Fermi liquid \cite{Rub99}.

We analyzed the electronic contribution for excitation energies
between 1.6 eV and 2.7 eV in B$_{1g}$ and B$_{2g}$ scattering
geometries. 
Fig.~\ref{wavelength} shows the Raman intensity in the B$_{1g}$ and
B$_{2g}$ symmetries for three overdoped Bi-2212 samples. In all cases
the overall
spectra at wavelengths $\lambda \neq 458$ nm have been multiplied by a
factor, ranging from 0.7 to 2, to adjust the intensity at 800-1000
cm$^{-1}$ to the spectra at $\lambda=458$ nm.
In both configurations we do not find any significant
variation of the lineshape of the spectra with excitation energy
at any doping level.
However, we observe a reduction of the overall intensity in the
B$_{2g}$ symmetry upon decreasing photon energies.
Hence, we cannot observe a specific variation of the pair breaking
feature with excitation energy neither in Bi-2212 nor in Y-123, except
for a small change in the overall cross section.

\begin{figure}
\centerline{\psfig{file=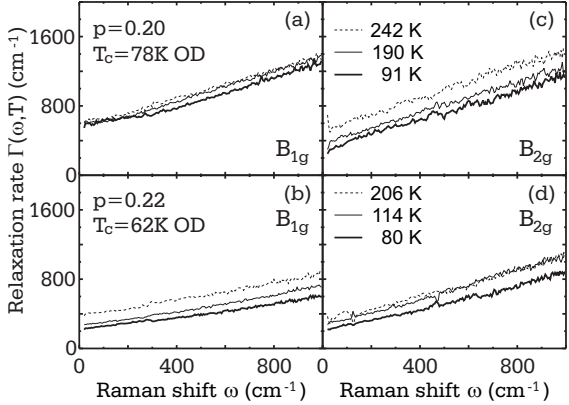,width=7.5cm,silent=}}
\caption{Relaxation rates in B$_{1g}$ ((a),(b)) and B$_{2g}$ ((c),(d))
symmetries at different temperatures for two overdoped samples.}
\label{rates}
\end{figure}

\section{Normal state}

The Raman continua in the normal state, $T > T_c$, (see e.g. Fig.
\ref{doping}) contain information on the dynamical two particle
lifetime $\tau$ for different regions in ${\bf k}$-space. In
Fig.~\ref{rates} we show the relaxation rates $\Gamma(\omega,T) = 1 /
\tau$ for two Bi-2212 samples with
$p$=0.20 and $p$=0.23, derived as described in \cite{Opel00}. The
contributions from the phonons have already been subtracted out.

Both in $B_{1g}$, Fig.~\ref{rates}(a,b), and $B_{2g}$,
Fig.~\ref{rates}(c,d), symmetries, the variation with frequency of the
quasiparticle relaxation rates show only little dependence on momentum
and doping. As compared to results at lower doping level
\cite{Opel00}, there is a tendency to a more quadratic frequency
dependence below approximately 400 cm$^{-1}$ in the B$_{1g}$ symmetry
possibly indicating more conventional quasiparticle dynamics.

The dependence of the dc limit of the relaxation rates
$\Gamma_0(T)=\Gamma(\omega \rightarrow 0,T)$ on
temperature evolves differently with doping in the two symmetries
analyzed. In particular, while $\Gamma_0(T)$ decreases with temperature at all
doping levels in the B$_{2g}$ symmetry, consistently with ordinary and
optical transport \cite{Tanner92,Ken93}, $\Gamma_0^{B_{1g}}(T)$ is
constant or slightly decreasing with $T$ for $p$=0.20 
and assumes the B$_{2g}$ behavior at $p\geq 0.22$. 
This disappearance of the anisotropy in the small doping range between
0.20 and 0.22 is directly visible in Fig.~\ref{rates}, where the $p=0.22$
sample shows a strongly suppressed relaxation rate in the B$_{1g}$ symmetry
(Fig.~\ref{rates}(b)).
We thus believe that the fundamental change in the 
carrier dynamics close to $(\pi,0)$ is an intrinsic property 
of the electron system itself and is closely 
related to correlation effects which are
found to fade away in this doping range in several other 
experiments \cite{Tall00}. This implies that the underlying 
interactions, ordering phenomena, and/or fluctuations 
must have some structure in momentum space.

\section{Conclusions}

We presented electronic Raman scattering results on Bi-2212. 
In the SC state, the B$_{1g}$ and B$_{2g}$ symmetry spectra are
characterized by very different doping dependences of the
superconductivity-induced features. While the B$_{2g}$ spectra are
consistent with a superconducting energy gap with $d$-wave symmetry
and a maximum $\Delta_0$ scaling with $T_c$, the B$_{1g}$ symmetry
requires further theoretical work.
In the normal state the quasiparticles at $(\pi,0)$ and at
$(\pi/2,\pi/2)$ are characterized by a qualitatively different doping
dependence of the relaxation rates. At high doping levels,
$p\geq 0.22$, the anisotropy observed for $p\leq 0.20$seems to disappear.

\vspace{0.5cm}

F.V. would like to thank the Gottlieb Daimler and Karl Benz Foundation
for financial support.

%%%%%%%%%%%%%%%%%%%%%%%%%%%%%%%%%%%%%%%%%%%%%%%%%%%%%%%
%Bibliografy
%%%%%%%%%%%%%%%%%%%%%%%%%%%%%%%%%%%%%%%%%%%%%%%%%%%%%%%

\end{document}